# Turbulence and Diffusion: Fossil Turbulence (MS 138)

Carl H. Gibson, Professor of Engineering Physics and Oceanography
Departments of MAE and Scripps Institution of Oceanography
University of California at San Diego, La Jolla, CA 92093-0411
cgibson@ucsd.edu, http://www-acs.ucsd.edu/~ir118

## Introduction

Fossil turbulence processes are central to turbulence, turbulent mixing, and turbulent diffusion in the ocean and atmosphere, in astrophysics and cosmology, and in other natural flows. However, because turbulence is often imprecisely defined, the distinct and crucial role of fossil turbulence may be overlooked. Turbulence occurs when inertial vortex forces $\vec{v} \times \vec{\omega}$ dominate all other forces for a range of length and time scales to produce rotational, eddy-like motions, where $\vec{v}$ is the velocity and $\vec{\omega}$ is the vorticity $\nabla \times \vec{v}$. Turbulence forms first at small viscous lengths with rapid overturn times; that is, at the Kolmogorov length scales $L_K \equiv (\nu^3/\varepsilon)^{1/4}$ and time scales $T_K \equiv (\nu/\varepsilon)^{1/2}$, where $\nu$ is the kinematic viscosity of the fluid and $\varepsilon$ is the viscous dissipation rate. Turbulence then cascades to larger scales where buoyancy forces cause fossilization of vertical motions at the Ozmidov scale $L_R \equiv (\varepsilon/N^3)^{1/2}$, where N is the ambient stratification frequency. Coriolis forces may cause fossilization at the Hopfinger scale $L_H \equiv (\varepsilon/\Omega^3)^{1/2}$, where $\Omega$ is the angular velocity. In the ocean and atmosphere this may occur at large horizontal scales, where $\Omega$ is the vertical component of the planetary angular velocity. Turbulence is defined as an eddy-like state of fluid motion where the inertial vortex forces of the eddies are larger than any other forces that tend to damp the eddies out. Fossil turbulence is defined as any fluctuation in a hydrophysical field such as temperature, salinity or vorticity that was produced by turbulence and persists after the fluid is no longer turbulent at the scale of the fluctuation. Fossil turbulence patches persist much longer than the turbulence events that produced them because they must mix the same velocity and scalar variances that existed in their original turbulence patches, but with smaller dissipation rates. They preserve information about the original turbulence, and complete the mixing and diffusion processes initiated by turbulence.

     The first printed reference to the concept of fossil turbulence was apparently when George Gamov suggested in 1954 that galaxies might be fossils of primordial turbulence produced by the Big Bang. Although it appears the primordial fluid at the time of galaxy formation was too viscous to be turbulent, the Gamov concept that fossilized hydrodynamic states might be reflected and preserved by parameters of the structures formed is correct. Persistent refractive index patches caused by mountain wakes in the stratified atmosphere and detected by radar were recognized as fossils of turbulence, causing organizers of the meeting to form a Fossil Turbulence working group for the 1969 Stockholm Colloquium on Spectra of Meteorological Variables, with a report by J. Woods (Editor), et al. in a special issue of Radio Science. An incorrect assumption of the report was that no universal fossil turbulence description is possible, and that all vertical velocity fluctuations vanish in

fossil turbulence. Woods showed that billows made visible by introducing dye in the interior of the stratified ocean formed highly persistent remnants of these turbulent events, as demonstrated by Thorpe in the laboratory using a tilt tube. Stewart termed patches of strong oceanic temperature microstructure measured without velocity microstructure from a submarine as "footprints of turbulence". A universal similarity theory of stratified fossil turbulence presented by Gibson in 1980 estimated universal constants of stratified fossil turbulence and introduced hydrodynamic phase diagrams as a method for classifying temperature and salinity microstructure patches in the ocean interior according to their hydrodynamic states (turbulent, active-fossil turbulence, completely fossil), and for extracting fossilized information about the previous turbulence and mixing. Most of the turbulent kinetic energy persists as universal, isotropic, saturated internal wave motions termed fossil vorticity turbulence. The theory is confirmed by measurements of Lozovatsky, Dillon, Gargett, Gregg, Osborne, Van Atta and others, and has been extended to describe fossilized turbulence produced by magnetic forces, self-gravitational forces and space-time inflation (see website http://www-acs.ucsd.edu/~ir118) in a new theory of gravitational structure formation by Gibson 1996-2000.

Fossil turbulence patches have often been mistaken for turbulence patches in oceanic turbulence sampling experiments that fail to account for the turbulent fossilization process. Such mistakes have led to large underestimates of the true average vertical turbulence flux rates, and are the source of the so-called "dark mixing" paradox of the deep ocean interior (a concept invented by Tom Dillon). Dark mixing is to the ocean as dark matter is to galaxies. Dark matter is unseen matter that must exist to prevent galaxies from flying apart by centrifugal forces. Dark mixing is mixing by turbulence events that must exist to explain why some layers in the ocean interior are well mixed, but have strong turbulent patches that are undetected except for their fossil turbulence remnants. Both the dark mixing and dark matter paradoxes are manifestations of the same problems; that is, extreme intermittency of nonlinear cascade processes over a wide range of values leading to extreme undersampling errors, and basic misunderstandings about the underlying irreversible fluid mechanics. In the ocean, observations of fossil turbulence patches of rare, powerful, but undetected turbulence events support the statistical evidence that bulk flow estimates of the vertical diffusivity in the deep main thermocline are correct rather than interpretations of sparse temperature dissipation rate measurements that claim large discrepancies but do no take either the extreme intermittency of in deep ocean layers or the fossil turbulence evidence into account.

Fossil turbulence signatures in hydrophysical fields preserve information about previous turbulence; for example, temperature fluctuations produced by turbulence for fluctuations at length scales where the turbulence has been damped by buoyancy are termed fossil-temperature-turbulence, just as skywriting rapidly becomes fossil-smoke-turbulence above the inversion layer. The larger the fossil temperature turbulence patch, the larger the viscous and temperature dissipation rates must have been in the original patch of turbulence. Fossil turbulence remnants are the footprints and scars of previous turbulent events, and redundantly preserve information about their origins in a wide variety of oceanic fields such as temperature, salinity, bubbles, and vorticity. The process of extracting information about previous turbulence and mixing from fossil turbulence is termed hydropaleontology.

Stratified and rotating fossils of turbulence are more persistent than their progenitor turbulence events because they possess almost the same velocity variance (kinetic energy) and scalar variance (potential entropy of mixing) as the original turbulent field, but have smaller viscous and scalar dissipation rates of these quantities than they had while they were fully turbulent just prior to the beginning of fossilization. The most powerful turbulence events produce the most persistent fossils, with persistence times proportional to the normalized Reynolds number $\varepsilon/\varepsilon_F$ and inversely proportional to the ambient stratification frequency N, where $\varepsilon_o \approx 3 L_T^2 N^3$ is the estimated dissipation rate at beginning fossilization and $\varepsilon_F = 30 \nu N^2$ is at complete fossilization, where $L_T$ is the maximum Thorpe overturning scale of the patch (Gibson, 1999). Oceanic fossil turbulence processes are more complex and important than fossil turbulence in non-stratified non-rotating flows where the only mechanism of fossilization is the viscous damping of turbulence before mixing is complete and no microstructure remains. Laboratory viscous fossil turbulence without stratification or rotation is thus mentioned in textbooks only as a curiosity of flow visualization, where eddy patterns of dye or smoke that appear to be turbulent are not because the turbulent fluid motions have been damped by viscosity. Buoyancy-fossils and rotation-fossils are difficult to study in the laboratory or in computer simulation because of the wide range of relevant length and time scales.

Most of the ocean's kinetic energy exists as fossil-vorticity-turbulence because its motions, driven by thermo-haline oceanic circulation, air-sea interaction of atmospheric motions and tidal forces of the sun and moon, are converted to turbulence energy at the top and bottom ocean surfaces by turbulence formation and its cascade to larger scales, and are not immediately or locally dissipated. Instead, oceanic turbulent kinetic energy and its induced scalar-potential-entropy is fossilized by buoyancy and Coriolis forces and distributed oceanwide by advection, leaving fossil-scalar-turbulence and fossil-vorticity-turbulence remnants in a variety of hydrodynamic states. These turbulence fossils move and interact with their environment and each other in the ocean interior by mechanisms that are poorly understood and hardly recognized. For example, a necessary stage of average double diffusive vertical fluxes in the ocean may be fluxes driven by double diffusive convection in the final stages of fossil-temperature-salinity-turbulence decay within fossil-temperature-salinity-turbulence patches. The turbulence event scrambles the pre-existing temperature and salinity fields to produce a stirred field in which the full range of possible double diffusive instabilities occur. These drive motions in the late stages of the turbulent fossil decay, leaving characteristic layered structures of salt fingering. Convective instabilities at low Rayleigh number, insufficient to drive turbulent motion, may also occur.

Powerful turbulence events produce fossils that radiate wave energy and trigger secondary turbulence events at their boundaries, this turbulence also fossilizes, and these fossils produce more turbulence. Turbulent mixing and diffusion is initiated by turbulence, but the final stages of the mixing and diffusion is completed only after the flow has become partially or completely fossilized. In practice, complete fossilization rarely occurs in the ocean because propagating internal waves will produce strong shears at the strong density gradients of fossil density turbulence boundaries through baroclinic torques $\nabla \rho \times \nabla p / \rho^2$, where $\rho$ is density and p is pressure.  Active turbulence that arises in this way from completely fossilized microstructure patches is known as zombie turbulence.

## History of fossil turbulence

The distinctive, eddy-like-patterns of turbulent motions are beautiful and easily recognized. Many ancient civilizations have woven them into their arts, religions and sciences.  The first attempts at hydropaleontology were applications of Kolmogorovian universal similarity theories of turbulence to cosmology by Gamov, Zel'dovich, Ozernoi and others of the Soviet school of turbulence.  Recent evidence from space telescopes suggests that primordial turbulence and density structure from the Big Bang were fossilized at $10^{-35}$ seconds by inflation of space beyond the length scales of causal connection ct of the fluctuations, where c is the speed of light and t is the time.  As in the ocean, this microstructure seeded the formation of all subsequent structures in a zombie-turbulence fossil-turbulence cascade that preserves evidence of the hydrophysical states of each stage of the process in various hydrodynamic fossils.

     The 1969 working group on fossil turbulence chaired by Woods examined patches of persistent refractive index fluctuations produced by turbulence in the stratified atmosphere and fluctuating temperature and dye patches produced by turbulence in the ocean interior which could not possibly be turbulent at the time of their detection. Woods and other scuba divers observed dye patch fossils of breaking internal waves from their turbulent beginning until they became motionless.  Stewart reported temperature microstructure patches from a submarine with and without measurable velocity fluctuations, indicating that those without must be fossilized.  Fossil turbulence patches produced in the stratified atmosphere by wind over mountains and wakes of other aircraft are dangerous to aircraft because of the long persistence times of stratified fossil vorticity turbulence patches. For example, a fossil turbulence patch sent passengers flying into the ceiling of a Boeing 737 at 24,000 feet the afternoon of September 3, 1999, bound from Los Angeles to San Francisco on United Flight 2036, injuring 15 of the 107 passengers and five crew aboard. Detectable refractive index fluctuations from billow turbulence events behind mountains are observed many km downstream by radar scattering measurements, long after all turbulence had been damped by buoyancy forces as shown by the layered anisotropy that develops in the radar returns.  Fossils of clear air turbulence CAT (referred to as "angels") were well known to radar operators long before the 1969 workshop.  The equivalent dominant turbulent patches of the deep ocean have not yet been detected in their actively turbulent states.

     A quantitative universal similarity theory of stratified fossil turbulence was published by Gibson in 1980 based on towed-body small-scale temperature measurements made in 1974 with John Schedvin in the upper ocean of the Flinders current off Australia in a US-Soviet intercomparison cruise with R. Ozmidov, V. Paka, I. Lozovatsky and V. Nabatov on the DMITRI MENDELEEV.  The data showed clear evidence of buoyancy effects causing departures from spectral forms of the Kolmogorov and Batchelor universal similarity theories of turbulence and turbulent mixing.  These departures were attributed to

stratified turbulence fossilization. The theory is illustrated in Figure 1 by the time evolution of velocity spectra $\phi_u$ and temperature spectra $\phi_T$ at five stages in a turbulent wake as the turbulence is fossilized by buoyancy forces. The integral of $\phi_u$ over wavenumber k is the velocity variance and the integral of $\phi_T$ is the temperature variance, where the Prandtl number $\nu/D$ is about 10 (corresponding to that of seawater) with D the thermal diffusivity. Spectra are multiplied by $k^2$ so their integrals represent velocity and temperature gradient variances and are proportional to the dissipation rates $\varepsilon$ and $\chi$, respectively.

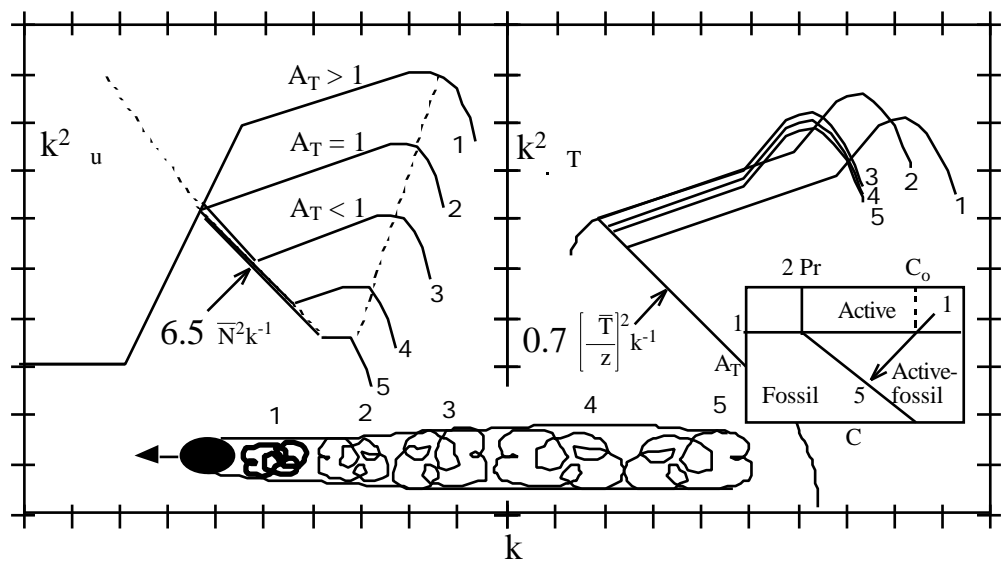

**Figure 1.** Dissipation spectra $k^2 \phi_u$ for velocity, left, and temperature $k^2 \phi_T$, right, for active turbulence in water as it fossilizes, represented by five stages of the temperature stratified turbulent wake at the bottom or, equivalently, at increasing times after the onset of turbulence in a patch (from Gibson 1999). Universal spectral forms for the saturated internal waves of the remnant fossil-vorticity-turbulence and fossil-temperature-turbulence microstructure patch are from Gibson (1980). The trajectory on an $A_T \equiv (\varepsilon/\varepsilon_o)^{1/2}$ versus Cox number C hydrodynamic phase diagram is shown in the right insert. Temperature dissipation rates $\chi$ provide a conservative estimate of $\varepsilon_o$ from the expression $\varepsilon_o \geq 13DCN^2$.

Stage 1 of the turbulence patch is fully turbulent with turbulent activity coefficient $A_T > 1$ ($\varepsilon > \varepsilon_o$, where $\varepsilon_o$ is the value at beginning of fossilization). Velocity dissipation spectra $k^2 \phi_u$ for stages 1-5 along the wake are shown on the left of Fig. 1. Each has integral $\varepsilon/3\nu$. Inertial subranges with slope +1/3 reflect Kolmogorov's second hypothesis for wavenumbers $k \geq 2\pi/L$ between the energy (or Obukhov) scale $L_O$ (small k) and the Kolmogorov scale $L_K$ (large k). The turbulence continues its cascade to larger $L_O$ scales by entrainment of external non-turbulent fluid until the increasing $L_O$ is matched by the decreasing Ozmidov scale $L_R = (\varepsilon/N^3)^{1/2}$ at the beginning of fossilization, where $\varepsilon = \varepsilon_o$, with spectral forms 2. The temperature dissipation spectrum $k^2 \phi_T$ increases in amplitude from stage 1 to 2 as vertical temperature differences are entrained over larger vertical scales, even though the velocity spectrum $k^2 \phi_u$ decreases, with increasing area $\chi/6D = C(\bar{T}/\partial z)^2/3$, where $\chi$ is the diffusive dissipation rate of temperature variance and Cox number C is the mean square over square mean temperature gradient ratio. The dramatic differences in spectral shapes between the velocity dissipation spectra and the temperature dissipation spectra in Fig. 1 reflect the theoretical result that without radiation, the kinetic energy of powerful fossilized turbulence events should persist as fossil-vorticity-turbulence for much longer periods than the temperature variance persists as fossil-temperature-turbulence. This is the basis of the $A_T \equiv (\varepsilon/\varepsilon_o)^{1/2}$ versus C hydrodynamic phase diagram shown at the bottom right of Fig. 1. Because $\chi$ and C are large in the temperature fossil, the velocity dissipation rate at beginning of fossilization $\varepsilon_o$ can be estimated from C using $\varepsilon_o \approx 3L_T^2 N^3 \approx$ 13DCN$^2$. The turbulent activity coefficient $A_T \equiv (\varepsilon/\varepsilon_o)^{1/2}$ is greater than 1 for stratified turbulence patches before fossilization, and less than 1 after fossilization begins. A particular patch decays along a straight line trajectory in the HPD until $\varepsilon = \varepsilon_F \approx 30 \nu N^2$ at complete fossilization. It can be shown that C averaged over a large horizontal layer in the stratified ocean for a long time period is a good measure of the turbulent heat flux divided by the molecular heat flux (Osborne and Cox 1972), with the vertical turbulent diffusivity K = DC. The motivation for most oceanic microstructure measurements is to estimate K through measurements of the average C for various oceanic layers. The problem is that C, $\varepsilon$, and $\chi$ in the ocean, atmosphere, and in all other such natural flows with wide cascade ranges, tend to be extremely intermittent.

## Intermittency of oceanic turbulence and mixing

Turbulence with the enormous range of length scales possible in oceanic layers is very intermittent in space and time. Sampling turbulence and turbulent mixing without recognizing this intermittency and without recognizing that most oceanic microstructure is fossilized or partially fossilized has led to misinterpretations and large errors in estimates of turbulent diffusion and mixing rates, especially in the deep ocean interior and in strong equatorial thermocline layers where intermittencies of $\varepsilon$ and $\chi$ are maximum. Dissipation rates $\varepsilon$ and $\chi$ are random variables produced by nonlinear cascades over a wide range of scales, resulting in lognormal probability density distributions and mean values larger than the mode values by factors in the range $10^2$ to $10^5$. Since the mode of a distribution is the most probable measured value, sparse microstructure studies will significantly underestimate mean $\varepsilon$ and $\chi$ values, as well as any vertical exchange coefficients and flux estimates of heat, mass, and momentum that are derived from these quantities. For lognormal random variables, $G_X = \exp[3\sigma^2_{\ln X}/2]$, where $\sigma^2_{\ln X}$ is the variance or intermittency factor of a lognormal random variable X. Intermittency factors $\sigma^2_{\ln \varepsilon}$ and $\sigma^2_{\ln \chi}$ in the ocean have been measured, and range from typical values of 5 at midlatitudes near the surface, to 6 or 7 in the deep ocean and at equatorial latitudes. Thus, probable undersampling errors for these quantities range from $G_\varepsilon$, $_\chi$ = 1800 to 36,000 in the ocean. For comparison, the intermittency factor $\sigma^2_{\ln \$}$ of the super-rich (upper 3%) in US personal income, which is close to lognormal, has been measured to be 4.3, giving a Gurvich number $G_\$$ of 600.





Figure 2 shows a lognormality plot of independent deep ocean samples of X = C averaged over 150 meters in the vertical. The axes are stretched so lognormal random variables fit a straight line.

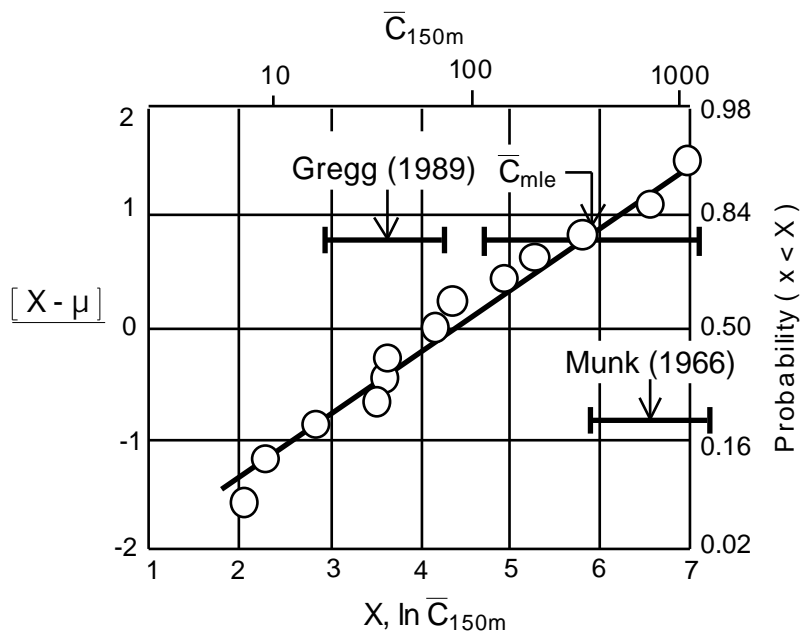

**Figure 2.** Normalized lognormal probability plot of dropsonde Cox number samples, averaged over 150 m in the vertical, in the depth range 75-1196 m, from Gregg (1977), expedition Tasaday 11, February, 1974. Estimated 95% confidence intervals are shown by horizontal bars, and compared to confidence intervals for C from the Gregg (1989) and Munk (1966) models for the thermocline, by Gibson (1991).

Fig. 2 shows the effects of undersampling errors due to intermittency in attempts to estimate the Cox number C of the stratified layers of the deep ocean, illustrating the "deep dark mixing paradox". In 1966 Munk estimating the vertical turbulent diffusivity of temperature in the deep Pacific Ocean below a kilometer depth should be $K = DC = 1$-$2$ $cm^2$ $s^{-1}$ with corresponding Cox number of 500-1100. Dropsonde temperature microstructure measurements find C values more than an order of magnitude less. However, the deep C averages are clearly lognormal, with maximum likelihood estimator $C_{mle}$ values in better agreement with the Munk range than with the microstructure range, as shown in Fig. 2. This matter is still controversial in the oceanographic literature, partly because the tests have either been carried out in shallow high latitude layers where there is no disagreement or in deep layers where adequate microstructure sampling is impossible. In the shallow main thermocline at 0.3 km, C values are less intermittent and K values are less by a factor of 30 by all methods, including tracer release studies. Thus the vertical heat flux is constant between 0.3 and 2 km at about 6 watts/$m^2$, consistent with computer models of planetary heat transfer.





Remarkably, the galactic dark matter paradox may exist for similar reasons. Gas emerging from the Big Bang plasma condenses to form widely separated objects with the mass of a small planet. These enter into nonlinear gravitational cascades to form stars a million times more massive, and their number density n becomes lognormal with Gurvich numbers $G_n$ near $10^6$. Only about 1 in 30 succeed. The rest are now dark and frozen, thirty million per star in a galaxy: the dark matter. Microlensing surveys have failed to detect these objects, and have excluded their existence assuming a uniform pdf rather than the expected intermittent lognormal probability density function for n that explains this incorrect interpretation (see http://xxx.lanl.gov/abs/astro-ph/9908335).

## Turbulence and fossil turbulence definitions

Turbulence is defined as a rotational, eddy-like state of fluid motion where the inertial-vortex forces of the eddies are larger than any of the other forces which tend to damp the eddies out. The inertial-vortex force $\vec{F}_I = \vec{v} \times \vec{\omega}$ produces turbulence, and appears in the Newtonian momentum conservation equations,

$$\frac{\partial \vec{v}}{\partial t} = -\nabla B + \vec{v} \times \vec{\omega} + \vec{F}_\nu + \vec{F}_C + \vec{F}_B + \ldots \;;\; B = \frac{p}{\rho} + \frac{v^2}{2} + gz, \quad (1)$$

where $\vec{v}$ is the velocity field, $\vec{\omega} = \nabla \times \vec{v}$ is the vorticity, B is the Bernoulli group of mechanical energy terms, p is pressure, $\rho$ is density, g is gravity, z is up, $\vec{F}_\nu = \nu \nabla^2 \vec{v}$ is the viscous force, $\nu$ is the kinematic viscosity, $\vec{F}_C = 2\vec{v} \times \vec{\Omega}$ is the Coriolis force, and $\vec{F}_B = N^2 L$ is the buoyancy force when the buoyancy frequency N is averaged over the largest vertical scale L of the turbulence event (other forces are neglected). The growth of turbulence is driven by $\vec{F}_I$ forces at all scales of the turbulent fluid. Irrotational flows (those with $\vec{\omega} = 0$) are nonturbulent by definition, but supply the kinetic energy of turbulence because the turbulent fluid induces a nonturbulent cascade of the irrotational fluid from large to small scales by sucking irrotational fluid into the interstices between the growing turbulence domains. In turbulent flows, viscous and inertial-vortex ($\vec{v} \times \vec{\omega}$) forces are equal at a universal critical Reynolds number $vx/\nu \approx 100$ for separation distances $x \approx 10\, L_K$, where $L_K$ is the Kolmogorov length scale

$$L_K \equiv \left[\frac{\nu^3}{\varepsilon}\right]^{1/4} \quad (2)$$

and $\varepsilon$ is the viscous dissipation rate per unit mass.

Fossil turbulence is defined as a fluctuation in any hydrophysical field produced by turbulence that persists after the fluid is no longer turbulent at the scale of the fluctuation. Examples of fossil turbulence are jet contrails, skywriting, remnants of cold milk poured rapidly into hot coffee, and patches of ocean temperature microstructure observed with little or no velocity microstructure existing within the patches. The best known fossil turbulence parameter in the ocean is the mixed layer depth, which persists long after it was produced and the turbulence has been damped. Buoyancy forces match inertial-vortex forces in a turbulent flow at the Ozmidov scale

$$L_R \equiv \left[\frac{\varepsilon}{N^3}\right]^{1/2}, \quad (3)$$





where the intrinsic frequency N of a stratified fluid is

$$N = \left[-\frac{g}{\rho}\frac{\partial \rho}{\partial z}\right]^{1/2}, \quad (4)$$

for the ambient stably stratified fluid affecting the turbulence. Coriolis forces match inertial-vortex forces at the Hopfinger scale

$$L_H = \left[\frac{\varepsilon}{\Omega^3}\right]^{1/4}, \quad (5)$$

where $\Omega$ is the angular velocity of the rotating coordinate system.

Taking the curl of Equation (1) for a stratified fluid gives the vorticity conservation equation

$$\frac{\partial \vec{\omega}}{\partial t} + \vec{v}\cdot\nabla\vec{\omega} = \vec{\omega}\cdot\overleftrightarrow{e} + \frac{\nabla\rho \times \nabla p}{\rho^2} + \nu\nabla^2\vec{\omega}, \quad (6)$$

where the vorticity of fluid particles (on the left side) tends to increase from vortex line stretching by the rate of strain tensor $\overleftrightarrow{e}$ (the first term on the right), baroclinic torques on strongly tilted strong density gradient surfaces (the second term), and decreases by viscous diffusion (the third term). From Equation (6) we see that turbulence events in stably stratified natural fluids are most likely to occur where density gradients are large and tilted for long time periods; for example on fronts, because this is where most of the vorticity is produced.

## Formation and detection of stratified fossil turbulence

Figure 3 shows a sequence of events for turbulence and fossil turbulence formation in the interior of the stratified ocean where vorticity is produced by tilting a density surface, starting from a state of rest. The longer the density surface is tilted the more kinetic energy is stored in the resulting boundary layers. The boundary layers formed on both sides of the tilted surface will become turbulent when a critical Reynolds number is reached. This occurs when the boundary layer thickness is 5-10 Kolmogorov scales based on the viscous dissipation rate of the laminar boundary layer, top right Fig. 3. The turbulence sharpens the density gradient, keeping the local Froude number $F(z) = u(z)/zN(z)$ less that $Fr_{crit}$. Billows form when $Fr(z)$ exceeds about 2, bottom left, and the turbulent burst occurs, absorbing all the kinetic energy stored on the tilted density layer. Fossilization begins when buoyancy forces match the inertial vortex forces of the turbulence, at a vertical size of about 0.6 $L_R$, as shown at bottom center, with viscous dissipation rate $\varepsilon_o$. The dissipation rate monotonically decreases with time during the process so $L_R$ decreases as the vertical patch size $L_P$ increases. The fossil turbulence patch does not collapse, even though the interior turbulent motions decrease in their vertical extent. Therefore, the patch size $L_P$ preserves information about the Ozmidov scale $L_{R_O}$ at the beginning of fossilization when the viscous dissipation rate $\varepsilon$ was $\varepsilon_o$. Thus, from Equation (3) and $L_{R_O} = \sqrt{3} L_P$ we have the expression

$$\varepsilon_o = 3L_P^2 N^3, \quad (7)$$

from which we can estimate $\varepsilon_o$ from measurements of $L_P$ and $N$ long after the turbulence event. Because the saturated internal waves of fossil vorticity turbulence have frequency $N$, they propagate vertically, and produce secondary turbulence events above and below the fossil turbulence patch, as shown in the bottom right of Fig. 3. Secondary turbulent events also form on the strong density gradients formed at the top and bottom of the fossil, because these gradient surfaces are likely to be strongly tilted.

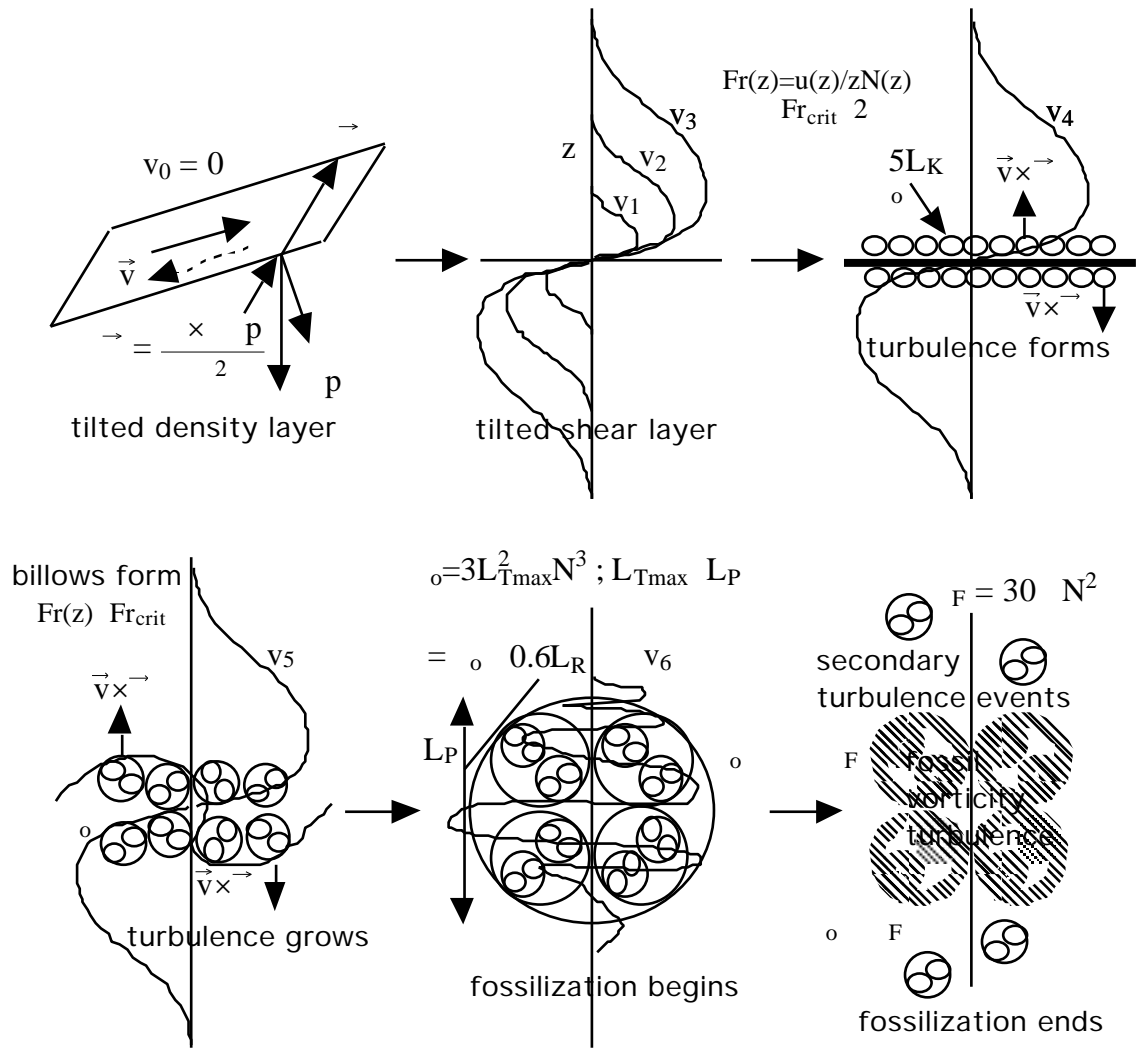

**Figure 3.** Fossil turbulence formation on a suddenly tilted density layer in a stagnant stratified fluid. Baroclinic torques cause a buildup of vorticity on the density surface at the top left, causing laminar boundary layers $v_1$ to $v_3$, top center. These become turbulent at 5-10 times the Kolmogorov scale $L_K$ as shown at top right. Billows form and the density interface is broken at the scale of the turbulent boundary layer corresponding to $Fr(z) \approx Fr_{crit} \approx 2$, bottom left. Turbulence cascades to larger vertical scales limited by approximately 0.6 times the Ozmidov scale $L_R$ where fossilization begins, bottom center, leaving a fossil vorticity turbulence remnant, which decays by viscous dissipation and vertical radiation of internal waves that may form secondary

turbulence events, bottom right. Microstructure patches are actively turbulent if $\varepsilon \geq \varepsilon_o$, partially fossilized for $\varepsilon_o \geq \varepsilon \geq \varepsilon_F$, and fossil for $\varepsilon \leq \varepsilon_F$.

Motions of the ocean are inhibited in the vertical direction by gravitational forces, so that the turbulence and fossil vorticity turbulence kinetic energy is mostly in the horizontal direction. In Equation (5), $\Omega$ is the vertical component of the Earth's angular velocity and approaches zero at the equator since $\Omega = \Omega \sin \phi$, where $\phi$ is the latitude. Large scale winds and currents develop at equatorial latitudes because they are unchecked by Coriolis forces. These break up into horizontal turbulence which can also cascade to large scales before fossilization by Coriolis forces at $L_H$ scales. Ozmidov and Hopfinger scales for the dominant turbulent events of particular layers, times, and regions of the ocean cover a wide range, with typical maximum values $L_R$ = 3-100 m, and $L_H$ = 30-500 km occurring where $\varepsilon$ is large and N and $\Omega$ are small.

## Quantitative Methods

A patch of temperature, salinity, or density microstructure is classified according to its hydrodynamic state by means of hydrodynamic phase diagrams (HPDs), one of which is shown in Fig. 1 (insert), which compare parameters of the patch to critical values. For the patch to be fully turbulent, both the Froude number Fr = U/NL and the Reynolds number Re = UL/$\nu$ must be larger that critical values from our definition of turbulence. If both are subcritical the patch is classified as completely fossilized. Most oceanic microstructure patches are found in an intermediate state, termed partially fossilized, where Fr is subcritical and Re is supercritical. This means that the largest turbulent eddies have been converted to saturated internal waves, but smaller scale eddies exist that are still overturning and fully turbulent. A variety of HPDs have been constructed as fossil turbulence theory has evolved, but all have active, active-fossil, and fully fossil quadrants. Figure 4 shows an HPD applied to turbulence-fossil turbulence-phytoplankton growth interaction.

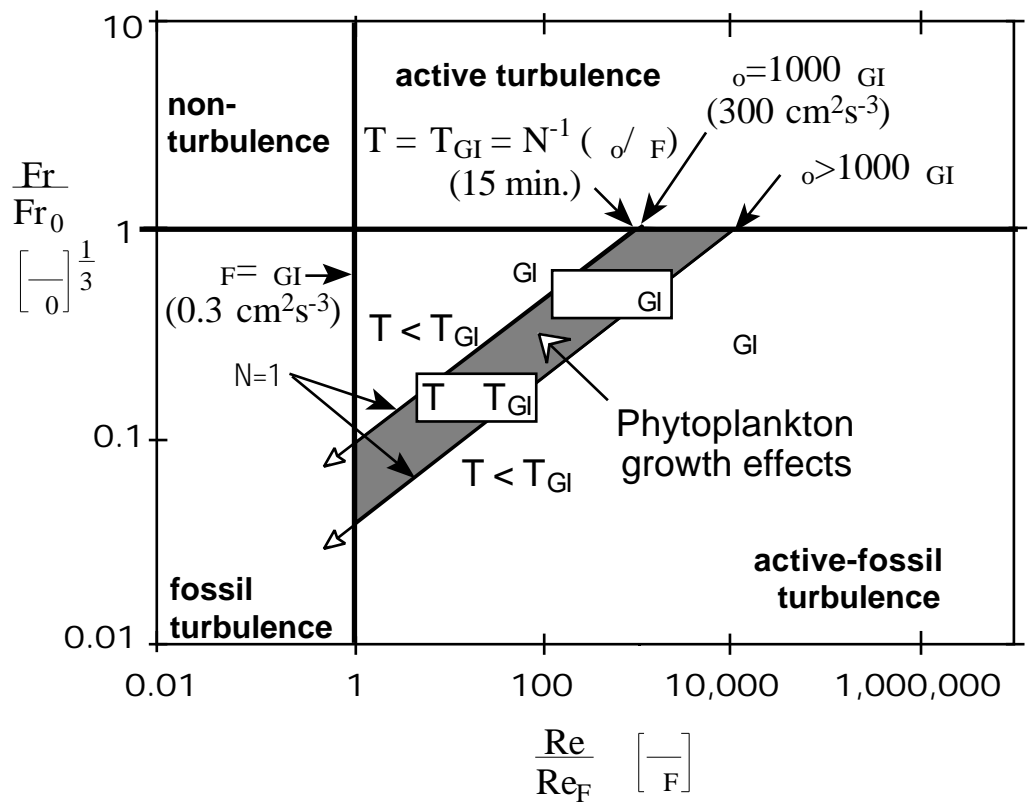

**Fig. 4** Hydrodynamic phase diagram showing the domain of phytoplankton growth effects corresponding to measured values of $\varepsilon_{GI} = 0.3$ cm$^2$ s$^{-3}$ and $T_{GI} = 15$ minutes (from Thomas, Tynan and Gibson 1997).

Growth rates of various phytoplankton species are extremely sensitive to both turbulence and the duration of the turbulence. Laboratory experiments reveal that red tide dinoflagellates have two thresholds for growth inhibition; dissipation rate $\varepsilon_{GI}$ for turbulence, and $T \geq T_{GI}$ for the duration T of turbulence with $\varepsilon \geq \varepsilon_{GI}$ (apparently to detect fossil turbulence from its greater persistence). If the dissipation rate $\varepsilon$ exceeds about $\varepsilon_{GI} = 0.3$ cm$^2$ s$^{-3}$ for more than $T = 15$ minutes a day for such microscopic swimmers they die in a few days. Shorter duration turbulence events are ignored, no matter how powerful. Diatom growth in the laboratory and field reacts positively to turbulence events with more than several minutes persistence. The hypothesis matching this behavior is that both classes of species have evolved methods of hydrodynamic pattern recognition so that they can maximize their chances of survival with respect to their swimming abilities.

Dinoflagellate red tides occur when nutrient rich upper layers of the sea experience several days of sun with weak winds and waves so that they become strongly stratified. The diatoms settle out of the light zone so that the dinoflagellates can bloom. However, when waves appear with sufficient strength to break and mix the surface layer, this may be detected by the phytoplankton from the long persistence time $T > T_{GI}$ of the fossil-vorticity-turbulence patches produced. It is supposed that phytoplankton species adjust their growth rates in anticipation of an upcoming sea state change from strongly stratified to well mixed according to their swimming abilities. The expression derived by the author relating the time of persistence of the turbulence patch before it becomes completely fossilized T = $N^{-1}$ / $_F$ to the Reynolds number ratio Re/Re$_F$ = / $_F$ = /30 $N^2$ is given at the top of Fig. 4. The shaded gray zone of the HPD in Fig. 4 shows estimates of N and that would inhibit growth of a particular red tide dinoflagellate species with known $_{GI}$ and T$_{GI}$.

## Further reading

References are in the Los Alamos E-print archive http://xxx.lanl.gov, and in website http://www-acs.ucsd.edu/~ir118.

Gibson, C. H., "Fossil Turbulence Revisited", Journal of Marine Systems, vol. 21, nos. 1-4, (1999) 147-167, astro-ph/9904237.

Gibson, C. H., "Turbulence in the ocean, atmosphere, galaxy, and universe," Applied Mechanics Reviews, 49:5, (1996) 299-315, astro-ph/9904260.

Gibson, C. H., "Kolmogorov Similarity Hypotheses for Scalar Fields: Sampling Intermittent Turbulent Mixing in the Ocean and Galaxy", in *Turbulence and stochastic processes: Kolmogorov's ideas 50 years on*, Proceedings of the Royal Society London, Ser. A, V 434 (N 1890) (1991), 149-164, astro-ph/9904269.